\documentclass{ws-p8-50x6-00}

\begin{document}

\title{Randomness in nuclei and in the quark-gluon plasma}

\author{ A. De Pace and A. Molinari }

\address{ Dipartimento di Fisica Teorica dell'Universit\`a di Torino and \\
 Istituto Nazionale di Fisica Nucleare, Sezione di Torino, \\
 via P.Giuria 1, I-10125 Torino, Italy }

\maketitle

\abstracts{ The issue of averaging randomness is addressed, mostly in nuclear
  physics, but shortly also in QCD. The Feshbach approach, so successful in
  dealing with the continuum spectrum of the atomic nuclei ({\em optical
  model}), is extended to encompass bound states as well ({\em shell model}).
  Its relationship with the random-matrix theory is discussed and the bearing
  of the latter on QCD, especially in connection with the spectrum of the Dirac
  operator, is briefly touched upon. Finally the question of whether Feshbach's
  theory can cope with the averaging required by QCD is considered. }

\section{The random-matrix theory and nuclear applications}
\label{sec:rm}

The random-matrix theory (RMT) is built out of any stochastic modelling of the
matrix representation of a Hamiltonian. It was introduced half-a-century ago by
Wigner, who considered an ensemble of Hamiltonians, different but sharing some
symmetry, having independent random variables as elements and searched for
those properties common to (nearly) all the members of the ensemble.
For this purpose he introduced the probability of the occurrence of a given
matrix $H$ in the ensemble by multiplying a weighting function $P_{N\beta}(H)$,
--- $N$ being the Hilbert space dimension and $\beta$ an index specifying the
symmetry of the ensemble, --- for the differentials of all the elements of the
matrix representing $H$. 

Wigner focused on matrices with random elements having a gaussian distribution
(Gaussian statistical hypothesis). The weighting function of a Hamiltonian with
$N$ quantum states is then 
\begin{equation}
\label{eq:PNb}
  P_{N\beta}(H) = e^{-\beta N \rm{tr}H^2/\lambda^2}, 
\end{equation}
$\lambda$ being some $N$-independent constant.

Concerning the symmetry, in the framework of the quantum Gaussian random-matrix
theory (GRMT) one can define three different ensembles:
\begin{itemize}
\item[i)] the ensemble of the real symmetric matrices, describing time-reversal
  and rotationally invariant systems, referred to as GOE (Gaussian orthogonal
  ensemble) and corresponding to $\beta=1$;
\item[ii)] the ensemble of Hermitian matrices, describing systems violating
  time-reversal invariance, as, e.~g., a nucleus in an external magnetic field,
  referred to as GUE (Gaussian unitary ensemble) and characterized by
  $\beta=2$; 
\item[iii)] the ensemble of the matrices that are linear combinations of the
  2x2 unit matrix and the Pauli matrices, namely
\begin{equation}
\label{eq:Hnm}
  H_{nm} = H^{(0)}_{nm} {\bf 1}_2 -i \sum_{j=1}^3 H^{(j)}_{nm} \sigma_j ,
\end{equation}
where $H^{(0)}_{nm}$ and $H^{(j)}_{nm}$ are real matrices, symmetric and
antisymmetric, respectively. The matrices (\ref{eq:Hnm}) describe systems with
half-integer spin, time-reversal, but not rotationally, invariant and the
associated ensemble is referred to as GSE (Gaussian symplectic ensemble), wich
corresponds to the choice $\beta=4$.
\end{itemize}
The elements of the matrices belonging to the ensembles GOE, GUE and GSE are
real, complex and quaternion numbers, respectively.

How to derive predictions on observables and how to compare these with the data
in the GRMT framework are questions recently reviewed by Weidenmueller {\em
et al.}.~\cite{Guh98} These authors emphasize that predictions having universal
validity, i.~e. unrelated to the specific nature of the system under
investigation, concern the fluctuations around mean values.
Typical in this connection is the energy spectrum of a complex system with its
mean level spacing $\Delta(\epsilon)$ and the related fluctuations: In the 
GRMT the latter are universal, the energies of the individual systems are not.

\begin{figure}[t]
\begin{center}
\epsfxsize=15pc
\epsfbox{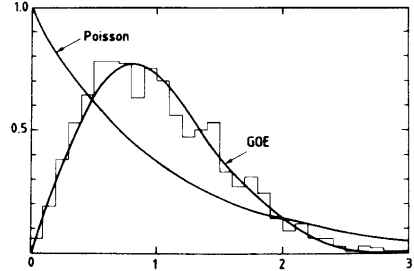}
\caption{ Nearest-neighbor spacing distribution versus the variable $s$ defined
  in the text. The GRMT prediction (denoted GOE) is compared to the data
  (histogram). Also shown the result obtained with a Poisson distribution.
  (Taken from Ref.~\protect\cite{Guh98}).
\label{fig:spacing}}
\end{center}
\end{figure}

More generally, are the local correlations among the eigenvalues (and the
eigenvectors as well) that are accounted for by the GRMT, using
$\Delta(\epsilon)$  as an input to fix the parameter $N/\lambda^2$ in
(\ref{eq:PNb}). As it is well-known, an amazing example in this connection is
offered by the short-range fluctuations in the spectra of the atomic nuclei
near the neutron emission threshold ($\sim8$~MeV). 
Indeed, given $s=S/\Delta(\epsilon)$ where
$S$ denotes the actual level spacing, the data of the nearest neighbour
spacing distributions of the nuclear levels, $p(s)$, are strikingly in accord
with the GRMT predictions as shown in Fig.~\ref{fig:spacing}.

It is of importance to realize that the strong repulsion displayed by the
nuclear levels, when their excitation energy is sufficiently large and their
distance sufficiently small, in the GRMT is engrained in Eq.~(\ref{eq:PNb}).
Indeed, if the eigenvalues $E_1$, $E_2$, $...E_N$ and the eigenvectors of the
matrices of the ensemble are chosen as new independent variables, then
(\ref{eq:PNb}) factorizes in two terms: One only eigenvalue, the other only
eigenvector dependent. The former reads
\begin{equation}
\label{eq:PNbE}
  P_{N\beta}(E_1..E_N) = \prod_{m>n} |E_m-E_n|^\beta \prod_{l=1}^N dE_l 
\end{equation}
and clearly displays the levels repulsion in the Vandermonde determinant that
multiplies $dE_1..dE_N$.

Actually, there is still a long way to go from (\ref{eq:PNbE}) to $p(s)$.
It is remarkable that Wigner was able to cross it with his famous ansatz
\begin{equation}
\label{eq:ps}
  p(s) = a s e^{-b s^2} ,
\end{equation}
which turned out to be very close to the exact predictions of the
GRMT.\cite{Gut90} Note that the gaussian fall-off of (\ref{eq:ps}) is not
related to the gaussian nature of the ensemble, but directly arises from the
Vandermonde determinant in (\ref{eq:PNbE}).

Applications of GRMT to QCD will be just touched upon in Sec.~\ref{sec:QCD}.

\section{Feshbach's theory}

\begin{figure}[t]
\begin{center}
\epsfxsize=13pc
\epsfbox{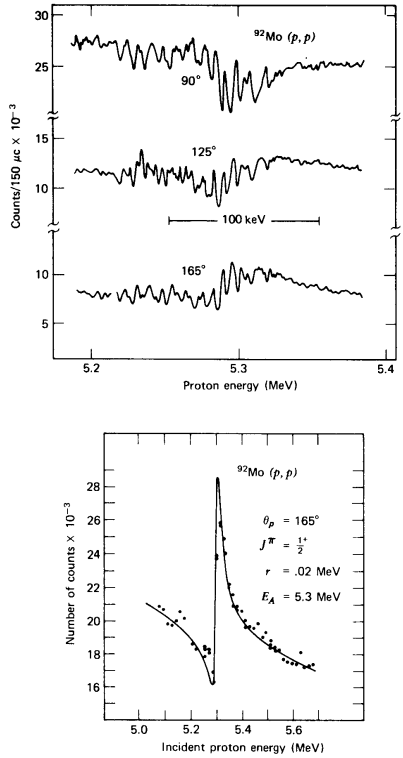}
\caption{(a) Elastic scattering of protons by $^{92}$Mo for energies close to
  the $s$-wave isobar analog resonance; (b) what one obtains by averaging the
  data of (a). (Taken from Ref.~\protect\cite{deS74}).
\label{fig:scatt}}
\end{center}
\end{figure}
Feshbach's approach assumes the existence of randomness in complex systems
whose internal structure is revealed by their excitation spectrum.
On the other hand, in this view there are no systems completely ruled by
randomness. The latter is defined through the relation
\begin{equation}
\label{eq:sAB}
  \langle \sigma(A) \sigma(B) \rangle = \langle \sigma(A) \rangle
    \langle \sigma(B) \rangle ,
\end{equation}
$\sigma$ being some physical observable of the system, dependent upon a
parameter: The system is chaotic if the value of $\sigma$ in correspondence of
the value $A$ for the parameter is independent of the value $\sigma$ assumes
for any other value of the parameter, say $B$.

Feshbach's~\cite{Fes00} view is that systems totally ruled by randomness do not
exist or, equivalently, that coherent combinations of $\sigma$ can always be
found such to allow the existence of correlations where now, in contrast with
the discussion of Sec.~\ref{sec:rm}, the correlations are {\em dynamical}, not
{\em statistical}, they relate to {\em order}, not to {\em disorder}.

The question then arises: How these dynamical correlations can be unraveled?
The answer is: With appropriate averaging procedures.
Before formally defining the average, it is illuminating to experience how it 
works in a specific example: The elastic
scattering of protons from the nucleus $^{92}$Mo at various scattering angles
and in the energy range of a few MeV. In Fig.~\ref{fig:scatt}a we show the 
data obtained with a poor energy resolution: They indeed appear to display the
features of a random process (note that the values $A$ and $B$ of a generic
parameter in (\ref{eq:sAB}) here correspond to specific values of the energy).
But in Fig.~\ref{fig:scatt}b the same data (at $\theta=165^{\circ}$) are
shown as obtained with a high energy resolution. What appears is astonishing: A
perfect resonance typical of a highly correlated scattering process.

The Heisenberg uncertainty principle $\Delta E \, \Delta t \ge \hbar/2$,
provides the key to understand. Poor energy resolution means large $\Delta E$,
hence small $\Delta t$ or short time for the system to interact: The data
appear random. On the other hand, small $\Delta E$ entails large $\Delta t$,
hence a long time for the system to average out the fluctuations:
The results are then those of Fig.~\ref{fig:scatt}b.

The necessity of averaging procedures to properly deal with randomness is thus
clear. This program has been carried out by Feshbach\cite{Fes92} in his unified
theory of nuclear reactions. We here extend his formalism to the bound states
of many-body systems, focusing on the ground state, to complete Feshbach's
program and to illustrate how it works. For this purpose we start
by partitioning the Hilbert space associated to a many-body system in the $P$
and $Q$ sectors, respectively.

\subsection{ Theoretical framework }
\label{subsec:basic}

As it is well-known, the splitting of the Hilbert space induced by the
projection operators $P$ and $Q$ transforms the Schroedinger equation 
\begin{equation}
  \label{eq:Schr}
  H\psi = E\psi
\end{equation}
into the pair of coupled equations
  \label{eq:Schrpair}
\begin{eqnarray}
  (E-H_{PP})(P\psi) &=& H_{PQ}(Q\psi) \\
  (E-H_{QQ})(Q\psi) &=& H_{QP}(P\psi) ,
\end{eqnarray}
the meaning of the symbols being obvious. From the above the equation obeyed
solely by $(P\psi)$ is derived. It reads
\begin{equation}
  \label{eq:Hcaleq}
  {\cal H}(P\psi) = E (P\psi),
\end{equation}
the $P$-space Hamiltonian being
\begin{equation}
  \label{eq:Hcal}
  {\cal H} = H_{PP}+H_{PQ}
    \frac{1}{\left(\frac{\displaystyle 1}{\displaystyle e_Q}\right)^{-1}+
    W_{QQ}} H_{QP},
\end{equation}
with 
\begin{eqnarray}
  e_Q    &=& E - H_{QQ} - W_{QQ} \\
  W_{QQ} &=& H_{QP}\frac{1}{E-H_{PP}}H_{PQ}.
\end{eqnarray}
It is of significance that although Eq.~(\ref{eq:Hcaleq}) is not an eigenvalue
equation, since the energy $E$ also appears in the denominator of its right
hand side, yet its solutions only occur for those values of $E$ which are
eigenvalues of (\ref{eq:Schr}) as well; in other words, a one-to-one
correspondence between the values of $E$ allowed by (\ref{eq:Schr}) and
(\ref{eq:Hcaleq}) exists (see later for a further discussion of this point).
However, the solutions of (\ref{eq:Hcaleq}) in correspondence to the various
values of $E$ {\em are not} orthogonal.

Now, we assume the quantum deterministic aspect of nuclear dynamics to be 
embodied in the $P$-space, the chaotic one in the $Q$-space. Hence, the
strategy of averaging over the latter follows, although, admittedly, some
fuzziness does affect this partitioning. To set up the averaging procedure we
start by the recognition that the wave functions in the $Q$-space are rapidly
varying functions of the energy $E$, viewed as a parameter classifying their
ensemble. 

Accordingly, we average over this ensemble following the prescription
\begin{equation}
  \label{eq:aver}
  \langle f(E)\rangle = \int_{-\infty}^{\infty}dE\,\rho(E,\bar{E}_0,\epsilon)
    f(E), 
\end{equation}
$f$ being a generic function to be averaged over the variable $E$ with the
distribution $\rho(E,\bar{E}_0,\epsilon)$.
The latter depends, beyond $E$, also upon the value $\bar{E}_0$ around which
the average, --- taken over a range of $E$ essentially set by $\epsilon$, ---
is performed.
A distribution convenient for our purposes is
\begin{equation}
  \label{eq:rho}
  \rho(E,\bar{E}_0,\epsilon) = \frac{1}{2\pi i}
    \frac{e^{iE\eta}}{E-(\bar{E}_0-\epsilon)-i\eta} ,
\end{equation}
which is indeed correctly normalized being 
\begin{equation}
  \int_{-\infty}^{\infty}dE\,\rho(E,\bar{E}_0,\epsilon) = 1
\end{equation}
(one should let $\eta\to0^+$ after the integration has been performed).
Note that Eq.~(\ref{eq:rho}) extends in some sense the Lorentz distribution of
the optical model\cite{Fes92} to the situation of a zero width state.
Hence the present formalism is especially suited to deal with ground states, 
which are of course stable: We shall accordingly focus mainly on the latter in
the following. 

Now, in the $Q$-space the wave functions are found to
be\cite{Fes96,Car98,DeP99}  
\begin{equation}
  \label{eq:Qpsi}
  (Q\psi) = \frac{1}{e_Q}H_{QP}\psi_0,
\end{equation}
$\psi_0$ being an auxiliary function that in the end disappears from the
formalism. By averaging Eq.~(\ref{eq:Qpsi}) according to the prescriptions
(\ref{eq:aver}) and (\ref{eq:rho}), one then finds that the averaged
wave function of the nuclear ground state in the $P$-space (here denoted by the
angle brackets) obeys the equation
\begin{equation}
  \label{eq:Hbar0eq}
  \bar{\cal H}\langle P\psi\rangle = \bar{E}_0 \langle P\psi\rangle.
\end{equation}
In (\ref{eq:Hbar0eq}) $\bar{E}_0$ is the mean field energy and
\begin{equation}
  \label{eq:Hbar}
  \bar{\cal H} = H_{PP}+H_{PQ}\frac{1}{\left\langle\frac{\displaystyle 1}
    {\displaystyle e_Q}\right\rangle^{-1}+W_{QQ}(E=\bar{E}_0)} H_{QP}
\end{equation}
is the mean field Hamiltonian.
This can be further elaborated since the singularities of the operator $1/e_Q$
lie in the Im$E<0$ half-plane\cite{Bro67}. Accordingly, one gets 
\begin{eqnarray}
  \label{eq:eQaver}
  \langle\frac{1}{e_Q}\rangle &=& \frac{1}{2\pi i}\int_{-\infty}^{\infty}dE
    \frac{e^{iE\eta}}{E-(\bar{E}_0-\epsilon)-i\eta}
    \frac{1}{E-H_{QQ}-W_{QQ}(E)} \\
  &=& \frac{1}{\bar{E}_0-\epsilon-H_{QQ}-W_{QQ}(E=\bar{E}_0-\epsilon)}
    \approx \frac{1}{\bar{E}_0-\epsilon-H_{QQ}-W_{QQ}(E=\bar{E}_0)}, \nonumber 
\end{eqnarray}
the last passage holding if the energy dependence of the operator $W_{QQ}$ is 
mild and if the parameter $\epsilon$ is not too large (it should be not too 
small either, otherwise the energy averaging procedure becomes meaningless).

The insertion of (\ref{eq:eQaver}) into (\ref{eq:Hbar}) leads then to the
following useful alternative expression for the mean field Hamiltonian
\begin{equation}
  \label{eq:Hbarav}
  \bar{\cal H} = H_{PP} + V_{PQ} V_{QP} \frac{1}{\bar{E}_0-\epsilon-E} ,
\end{equation}
where the energy dependent operators
\begin{equation}
  V_{PQ} = H_{PQ}\sqrt{\frac{\bar{E}_0-\epsilon-E}{\bar{E}_0-\epsilon-H_{QQ}}}
  \label{eq:VPQ}
\end{equation}
and
\begin{equation}
  V_{QP} = \sqrt{\frac{\bar{E}_0-\epsilon-E}{\bar{E}_0-\epsilon-H_{QQ}}}H_{QP},
  \label{eq:VQP}
\end{equation}
represent the residual effective NN interaction.
The usefulness of the Eqs.~(\ref{eq:Hbarav}), (\ref{eq:VPQ}) and (\ref{eq:VQP})
was realized  
in Ref.~\cite{Kaw73}, where it was noticed that with their help the pair of
equations (\ref{eq:Schrpair}) can be recast, as far as $(P\psi)$ is concerned,
into the form
\begin{eqnarray}
  \label{eq:Schrava}
  (E-\bar{\cal H})(P\psi) &=& V_{PQ}(Q\psi) \\
  \label{eq:Schravb}
  (E-H_{QQ})(Q\psi)       &=& V_{QP}(P\psi) ,
\end{eqnarray}
which is suitable for expressing the mean field fluctuations (the ``error'').

Indeed, by exploiting the spectral decomposition of the operator 
$(E-\bar{\cal H})^{-1}$  in terms of the eigenfunctions $\phi_n$ of the mean
field Hamiltonian $\bar{\cal H}$,\footnote{We omit here the difficult proof
concerning the normalization and orthogonalization of the eigenfunctions
$|\phi_n\rangle$ of $\bar{\cal H}$.} one gets from Eq.~(\ref{eq:Schrava})
\begin{eqnarray}
  \label{eq:spect}
  |P\psi\rangle &=& \sum_n\frac{|\phi_n\rangle}{E-\bar{E}_n}
    \langle\phi_n|V_{PQ}|Q\psi\rangle \nonumber \\
  &=& |\phi_0\rangle\frac{\langle\phi_0|V_{PQ}|Q\psi\rangle}{E-\bar{E}_0} +
    \left(\frac{1}{E-\bar{\cal H}}\right)^\prime V_{PQ}|Q\psi\rangle,
\end{eqnarray}
which, upon left multiplication by $\langle\phi_0|$, yields
\begin{equation}
  \label{eq:phi0ppsi}
  \langle\phi_0|P\psi\rangle = 
    \frac{\langle\phi_0|V_{PQ}|Q\psi\rangle}{E-\bar{E}_0}.
\end{equation}
In the second term on the right hand side of Eq.~(\ref{eq:spect}), the prime
stands for the omission of the $n=0$ term in the spectral decomposition.

Next, the insertion of Eq.~(\ref{eq:spect}) into (\ref{eq:Schravb}) leads to
\begin{equation}
  \label{eq:Qpsipsi}
  |Q\psi\rangle =
   \frac{1}{E-h_{QQ}}V_{QP}|\phi_0\rangle\langle\phi_0|P\psi\rangle,
\end{equation}
where the operator
\begin{equation}
  \label{eq:hQQ}
  h_{QQ} = H_{QQ} + V_{QP}\left(\frac{1}{E-\bar{\cal H}}\right)^\prime V_{PQ}
\end{equation}
has been introduced. Finally, by combining (\ref{eq:phi0ppsi}) and
(\ref{eq:Qpsipsi}), one arrives at the equation
\begin{equation}
  \label{eq:E-E0}
  E-\bar{E}_0 = \langle\phi_0|V_{PQ}\frac{1}{E-h_{QQ}}V_{QP}|\phi_0\rangle,
\end{equation}
which is the basis for computing the mean field energy error (or the
fluctuations of the energy associated with randomness).

Although Eq.~(\ref{eq:E-E0}) is valid for any choice of the projectors $P$ and
$Q$, its use is in fact appropriate when the $P$-space is one-dimensional,
as it was indeed the case in Ref.~\cite{DeP99}, where this choice was made for
sake of simplicity. 
For a two-dimensional $P$-space, as we shall later see, one should rather 
single out two, rather then one,  terms in the spectral decomposition of the
operator $(E-\bar{\cal H})^{-1}$ on the right hand side of 
Eq.~(\ref{eq:spect}).

We refer the reader to Refs.~\cite{Fes96,Car98,DeP99} for a discussion on how
the average of the square of Eq.~(\ref{eq:E-E0}) is actually computed
(the average of (\ref{eq:E-E0}) of course should vanish) and on how
the complexity expansion is organized. Here, we confine
ourselves to the leading term of this fast converging expansion.

\subsection{ Energy averaging }
\label{subsec:eneave}

To understand better the significance of the energy averaging distribution 
(\ref{eq:rho}) we show how it works in the simple cases of a 
bi-dimensional (A) and of a tri-dimensional (B) Hilbert space.

\begin{center}
{\bf A. Bi-dimensional Hilbert space }
\end{center}

Let $|\chi_1\rangle$ and $|\chi_2\rangle$ be the two normalized states
spanning the space. Here the only possible choice for the projectors clearly is
\begin{equation}
  P\equiv|\chi_1\rangle\langle\chi_1| \qquad{\rm and}\qquad
    Q\equiv|\chi_2\rangle\langle\chi_2|.
\end{equation}
Then, by expanding the operator $1/(E-H_{QQ})$, Eq.~(\ref{eq:Hcaleq}) can be 
recast as follows
\begin{equation}
  \left[E-|\chi_1\rangle a_{11}\langle\chi_1|
    -|\chi_1\rangle a_{12}\langle\chi_2|\frac{1}{E}\sum_{n=0}^{\infty}
    \left(\frac{a_{22}}{E}\right)^n(|\chi_2\rangle\langle\chi_2|)^n
    |\chi_2\rangle a_{12}^*\langle\chi_1|\right]|P\psi\rangle=0,
\end{equation}
which, upon multiplying from the left by $\langle\chi_1|$ and exploiting the
idempotency of $|\chi_2\rangle\langle\chi_2|$, simplifies to
\begin{equation}
  \left[E-a_{11}-\frac{|a_{12}|^2}{E-a_{22}}\right]
    \langle\chi_1|P\psi\rangle = 0,
\end{equation}
where the shorthand notations
\begin{equation}
  a_{11}=\langle\chi_1|H|\chi_1\rangle,\quad
  a_{22}=\langle\chi_2|H|\chi_2\rangle \quad{\rm and}\quad
  a_{12}=\langle\chi_1|H|\chi_2\rangle
\end{equation}
have been introduced.
This equation is trivially solved yielding the eigenvalues
\begin{equation}
\label{eq:E+-A}
  E_{\pm} =
    \frac{1}{2}\left[a_{11}+a_{22}\pm\sqrt{(a_{11}-a_{22})^2+4|a_{12}|^2}
    \right],
\end{equation}
which coincide with those of $H$.
It helps notice that the eigenvalues (\ref{eq:E+-A}) are also found as 
intersections of the hyperbola 
\begin{equation}
  \label{eq:hyper}
  E = a_{11} + \frac{|a_{12}|^2}{\omega-a_{22}}
\end{equation}
with the straight line $E=\omega$.

Also the energy averaged Hamiltonian (\ref{eq:Hbarav}) can be expressed in the
basis spanned by $\chi_1$ and $\chi_2$ and one gets the mean field equation
\begin{equation}
  \left[\bar{E}-a_{11}-\frac{|a_{12}|^2}{\bar{E}-\epsilon-a_{22}}\right]
  \langle\chi_1|\big\langle P\psi\big\rangle \rangle = 0.
\end{equation}
The latter is again trivially solved yielding
\begin{equation}
  \label{eq:Ebar+-A}
  \bar{E}_{\pm} =\frac{1}{2}\left[a_{11}+a_{22}+\epsilon\pm
    \sqrt{(a_{11}-a_{22}-\epsilon)^2+4|a_{12}|^2}\right],
\end{equation}
which now corresponds to the intersections of the hyperbola (\ref{eq:hyper})
(with $\bar{E}$ replacing $E$) with the new straight line
$\bar{E}=\omega+\epsilon$. 

\begin{figure}[t]
\begin{center}
\epsfxsize=10pc
\epsfbox{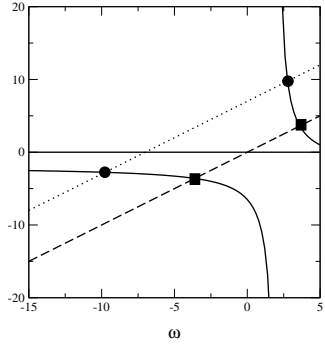}
\caption{ The eigenvalues of a bi-dimensional Hilbert space. The matrix
elements of the Hamiltonian are taken to be $a_{11}=-2$, $a_{22}=2$ and 
$a_{12}=3$, in arbitrary units.
The exact eigenvalues (squares) and the ones of the energy averaged Hamiltonian
$\bar{\cal H}$ (circles) are shown. They correspond to the intersections with
the straight line $E=\omega+\epsilon$: The former with $\epsilon=0$, the latter
with $\epsilon=7$. The stability of the lowest eigenvalue and the upward shift
of the highest one are clearly apparent. 
\label{fig:hyperbola1}}
\end{center}
\end{figure}

From Fig.~\ref{fig:hyperbola1}, where the solutions $E_{\pm}$ and
$\bar{E}_{\pm}$ are graphically displayed, it clearly appears that, while
$\bar{E}_{-}\cong E_{-}$, the solution $\bar{E}_{+}$ is much larger than
$E_{+}$, the more so the greater $\epsilon$ is.
It is thus clear that the averaging distribution (\ref{eq:rho}), while 
mildly affecting the eigenvalue of $H$ lying in the $P$-space, drives away the
one lying in the $Q$-space.

\begin{figure}[t]
\begin{center}
\epsfxsize=10pc
\epsfbox{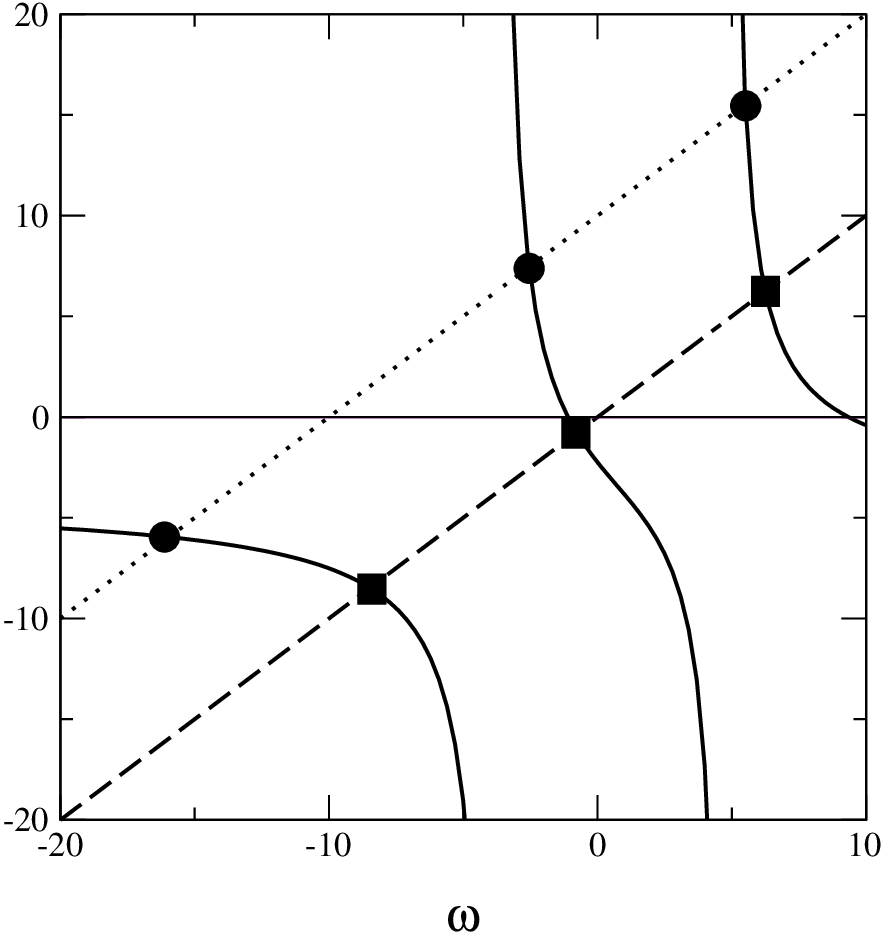}
\caption{ The eigenvalues of a tri-dimensional Hilbert space: The case of a
one-dimensional $P$-space. The following matrix elements of the Hamiltonian are
taken: $a_{11}=-4$, $a_{22}=-2$, $a_{33}=3$, $a_{12}=2$,
$a_{13}=-5$ and $a_{23}=3.5$, in arbitrary units.
The exact eigenvalues (squares) and the ones of the energy averaged Hamiltonian
$\bar{\cal H}$ (circles) are shown. They correspond to the intersections with
the straight line $E=\omega+\epsilon$: The former with $\epsilon=0$, the latter
with $\epsilon=10$. The stability of the $P$-space eigenvalue and the upward
shift of those belonging to the $Q$-space are clearly apparent. 
\label{fig:hyperbola2}}
\vskip 1cm
\epsfxsize=10pc
\epsfbox{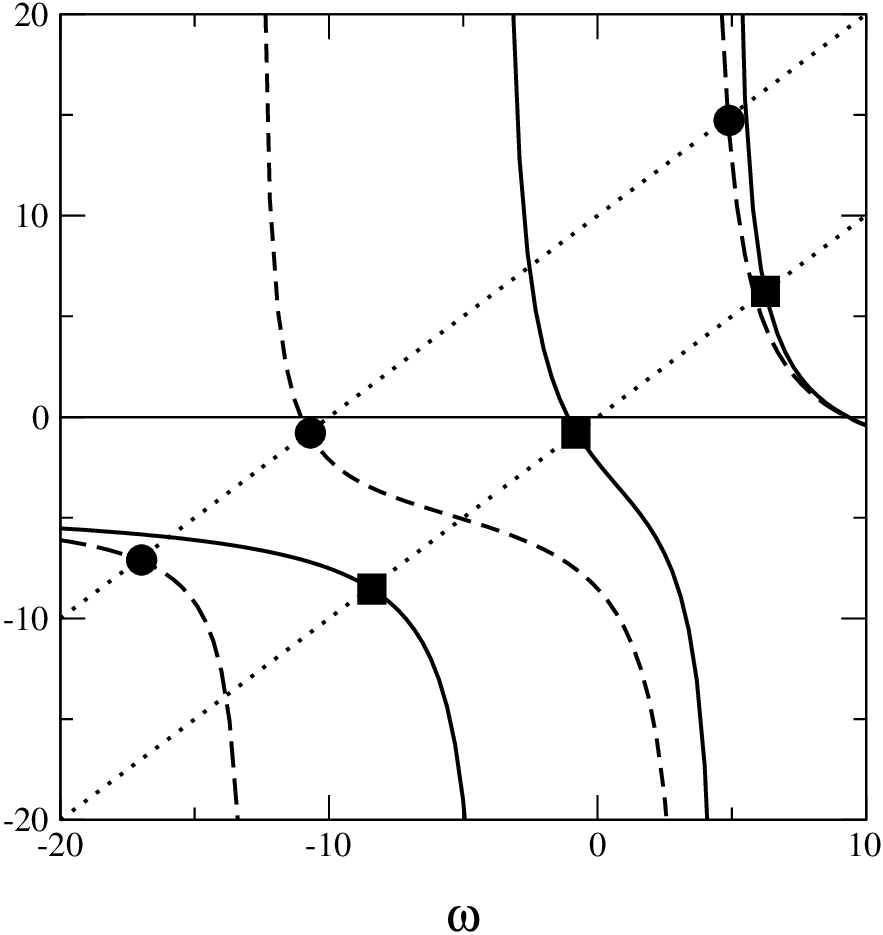}
\caption{ The eigenvalues of a tri-dimensional Hilbert space: The case of a 
bi-dimensional $P$-space. The matrix elements of the Hamiltonian are taken as 
in Fig.~\protect\ref{fig:hyperbola2}. The exact eigenvalues (squares) 
correspond to the intersections of the straight line $E=\omega$ with the
continuous curve.
The eigenvalues of the energy averaged Hamiltonian $\bar{\cal H}$ correspond to
the intersections of the straight line $\bar{E}=\omega+\epsilon$, with
$\epsilon=10$, with the dashed curves. Note the dependence upon $\epsilon$. 
The stability of the $P$-space eigenvalue and the upward shift of the one 
belonging to the $Q$-space are again clearly apparent. 
\label{fig:hyperbola3}}
\end{center}
\end{figure}

\begin{center}
{\bf B. Tri-dimensional Hilbert space }
\end{center}

The space is spanned by the normalized states $|\chi_1\rangle$,
$|\chi_2\rangle$ and $|\chi_3\rangle$. Now, two choices are possible for the
projectors, namely
  \label{eq:part1}
\begin{eqnarray}
  P &\equiv& |\chi_1\rangle\langle\chi_1| + |\chi_2\rangle\langle\chi_2| \\
  Q &\equiv& |\chi_3\rangle\langle\chi_3|
\end{eqnarray}
and
  \label{eq:part2}
\begin{eqnarray}
  P &\equiv& |\chi_1\rangle\langle\chi_1| \\
  \label{eq:part2b}
  Q &\equiv& |\chi_2\rangle\langle\chi_2| + |\chi_3\rangle\langle\chi_3|.
\end{eqnarray}
In both cases, Eq.~(\ref{eq:Hcaleq}) can be recast as follows
\begin{equation}
  \label{eq:Ecub}
  (E-a_{11})(E-a_{22})(E-a_{33})-|a_{12}|^2(E-a_{33})-|a_{13}|^2(E-a_{22})
    -|a_{23}|^2(E-a_{11})=0,
\end{equation}
which is the cubic equation yielding the exact eigenvalues.
Note that Eq.~(\ref{eq:Ecub}) is easily obtained with the choice
(\ref{eq:part1}), because in this case the operator $(E-H_{QQ})^{-1}$ is
expanded in terms of the idempotent operator $|\chi_3\rangle\langle\chi_3|$.
Not so with the choice (\ref{eq:part2}), because now $(E-H_{QQ})^{-1}$ should
be expanded in terms of the operator (\ref{eq:part2b}), {\em which is not
idempotent}. Actually, the larger the powers of the latter are, the more
cumbersome they become. Yet, also in this case it can be proved that
Eq.~(\ref{eq:Ecub}) holds valid.

Let us now examine the solutions of Eq.~(\ref{eq:Hbarav}): As in the previous
bi-dimensional case it is convenient to display the solutions graphically.
For the partition (\ref{eq:part2}), one finds that they are given by
the intersections of the $\epsilon$-independent curve
\begin{equation}
  \label{eq:EbarB}
  \bar{E} = a_{11} + \frac{|a_{12}|^2(\omega-a_{33})}{{\cal D}_1(\omega)}
    + \frac{|a_{13}|^2(\omega-a_{22})}{{\cal D}(\omega)}
    + \frac{a_{12}a_{13}^{*}a_{23}}{{\cal D}(\omega)}
    + \frac{a_{12}^{*}a_{13}a_{23}^{*}}{{\cal D}(\omega)},
\end{equation}
where
\begin{equation}
  \label{eq:Dcal}
  {\cal D}(\omega) = (\omega-a_{22})(\omega-a_{33})-|a_{23}|^2,
\end{equation}
with the straight line $\bar{E}=\omega+\epsilon$, as displayed in
Fig.~\ref{fig:hyperbola2}, where the case $\epsilon=0$, --- which clearly
provides the exact eigenvalues $E_i$ of the Schroedinger equation, --- is also
shown. From the figure, it transparently appears that $\bar{E}_0\cong E_0$,
whereas $\bar{E}_1>>E_1$ and $\bar{E}_2>>E_2$, the latter inequalities being
stronger when the parameter $\epsilon$ is large.

In the case of the partition (\ref{eq:part1}), the solutions are given by
the intersections of the curve obtained replacing $a_{22}\to a_{22}-\epsilon$ 
in Eqs.~(\ref{eq:EbarB}) and (\ref{eq:Dcal}) with the straight line
$\bar{E}=\omega+\epsilon$, as displayed in Fig.~\ref{fig:hyperbola3}.
We face here a new situation, since now not only the straight line, but also
Eqs.~(\ref{eq:EbarB}) and (\ref{eq:Dcal}) are $\epsilon$-dependent.
Yet, one again sees that for $\epsilon=0$ one recovers the eigenvalues $E_i$,
whereas when $\epsilon\ne0$ the intercepts occur for $\bar{E}_0\cong E_0$ and
$\bar{E}_1\cong E_1$, but for $\bar{E}_2>>E_2$.
Hence, we conclude that the action of the averaging distribution (\ref{eq:rho})
affects very little the eigenvalues belonging to the $P$-space, while pushing
off the ones in the $Q$-space by an amount proportional to $\epsilon$.

\subsection{ The $P$-space }
\label{subsec:Pspace}

Having defined the energy averaging procedure, to get further it is necessary
to define the operators $P$ and $Q$. 
For this purpose the natural candidates as building
blocks of the $P$ operator appear to be the eigenstates $|\phi_n\rangle$ of the
mean field Hamiltonian (\ref{eq:Hbarav}), defined by the equation 
\begin{equation}
  \label{eq:Hbareq}
  \bar{\cal H} |\phi_n\rangle = \bar{E}_n |\phi_n\rangle.
\end{equation}
Their finding requires, however, the solution of a difficult self-consistency 
problem.
Hence, we make the simpler choice of viewing as building blocks of $P$ the
Hartree-Fock (HF) variational solutions, which are, e.~g., trivial in nuclear
matter, the system we shall consider.

For sake of illustration we start with a one-dimensional $P$-space by setting
\begin{equation}
  \label{eq:Pone}
  P = |\chi_{\rm HF}\rangle\langle\chi_{\rm HF}|,
\end{equation}
$|\chi_{\rm HF}\rangle$ being the HF ground state wave function of nuclear
matter (the Fermi sphere).
Then, on the basis of (\ref{eq:Pone}), one derives the mean
field equation
\begin{equation}
  \label{eq:E0barone}
  \bar{E}_0 = E_{\rm HF} + \frac{\beta^2}{\bar{E}_0-\epsilon-E},
\end{equation}
which relates the mean field ($\bar{E}_0$), the HF ($E_{\rm HF}$) and the
true ($E$) energies {\em per particle}, and the equation for the statistical
fluctuation of the energy
\begin{equation}
  \label{eq:fluctone}
  E-\bar{E}_0 = \pm\frac{1}{E-\bar{\epsilon}_2}\sqrt{\frac{2}{{\cal N}_2}}
  \beta^2, 
\end{equation}
where
\begin{equation}
  \label{eq:beta2}
  \beta^2 = 
  \sum_{\rm 2p-2h}|\langle\psi_{\rm 2p-2h}|V|\chi_{\rm HF}\rangle|^2,
\end{equation}
the bras $\langle\psi_{\rm 2p-2h}|$ representing the two-particle--two-holes
(2p-2h) states of nuclear matter, whose average energy {\em per particle} is
$\bar{\epsilon}_2$. 
Thus all the quantities appearing in Eqs.~(\ref{eq:E0barone}) and
(\ref{eq:fluctone}) are {\em per particle}, including the parameter $\epsilon$
and the residual effective interaction $V$ (which are accordingly divided 
by the nuclear mass number $A$).

\begin{figure}[t]
\begin{center}
\epsfxsize=15pc
\epsfbox{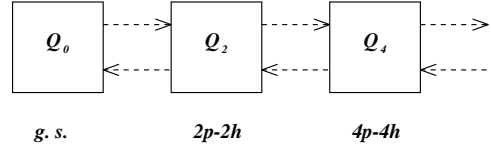}
\caption{The partition of the Hilbert space of nuclear matter in sets of
  increasing complexity. The first box on the left defines the $P$-space, the
  second one embodies the simplest states in the $Q$-space and so on.
\label{fig:boxes}}
\end{center}
\end{figure}

Note also that Eq.~(\ref{eq:fluctone}) gives the ``fluctuations'' of the mean
field energy in the first order of the complexity expansion, which is based on
an organization of the $Q$-space in blocks of excited states of increasing
complexity (see Fig.~\ref{fig:boxes}): Here, the contribution to the error only
arises from the sector of the $Q$-space set up with the 2p-2h excitations.
Moreover, although the states of the $Q$-space obey well-defined, coupled
differential equations (see Ref.~\cite{Car98}), we describe them with the HF 
multi-particle--multi-hole solutions, an approximation not impairing the 
orthogonality constraint $P\cdot Q=0$.

The sum in Eq.~(\ref{eq:beta2}) is performed over the ensemble of the 2p-2h
excited states lying in an appropriate energy range (in
Refs.~\cite{Car98,DeP99} taken to be fixed essentially by the parameter
$\epsilon$), whose number ${\cal N}_2$ can be computed using the Ericson's
formula\cite{Eri60} for the density of the spin $J$ $N$-particle--$N$-hole
nuclear states, namely
\begin{equation}
  \label{eq:rhoN}
  \rho^{(N)}_{ph}({\cal E},J) = \frac{g(g{\cal E})^{N-1}}{p!h!(N-1)!}
    \frac{2J+1}{\sqrt{8\pi}\sigma^3 N^{3/2}} \exp[-(2J+1)^2/(8N\sigma^2)] ,
\end{equation}
where
\begin{equation}
  \label{eq:gsigma}
  g = \frac{3}{2}\frac{A}{\epsilon_F} 
    \qquad {\rm and} \qquad
  \sigma^2 = {\cal F}\sqrt{\frac{\cal E}{a}}\frac{1}{\hbar^2},
\end{equation}
${\cal F}$ being the nuclear moment of inertia, $\epsilon_F$ the Fermi energy,
$a=A/8$ MeV$^{-1}$ and ${\cal E}$ the excitation energy of the system.

Now, Eqs.~(\ref{eq:E0barone}) and (\ref{eq:fluctone}), owing to the double
sign appearing in the latter, set up two systems, each one including two
equations, in two unknowns. Two options are then possible in selecting the
latter: One can choose either the ground state mean field and true energies per
particle, i.~e. $\bar{E}_0$ and $E$, --- assuming the matrix elements of the
residual effective interaction to be known, --- or $\bar{E}_0$ and
$\beta^2$, when $E$ is experimentally known, --- which is indeed the case in
nuclear matter.

We take the latter choice, requiring in addition the coincidence of the two
$\bar{E}_0$ obtained by solving the two systems separately.
Actually, and notably, both systems lead to the same formal expression for the
mean field energy per particle, namely
\begin{equation}
  \label{eq:E0barsol}
  \bar{E}_0 = \frac{1}{2} \left\{(E_{\rm HF}+E+\epsilon) -
    \sqrt{(E_{\rm HF}-E-\epsilon)^2+4\beta^2} \right\} ,
\end{equation}
which holds valid for $\epsilon<E_{\rm HF}-E$, the right hand side of this
inequality being positive because of the variational principle.
The above, when $\beta^2\to0$, yields
\begin{equation}
  \bar{E}_0 = E + \epsilon,
\end{equation}
in accord with (\ref{eq:VPQ}) and (\ref{eq:VQP}), but also with
(\ref{eq:fluctone}), which, for $V\to0$, gives $\bar{E}_0=E$: Indeed, when the
residual effective interaction vanishes no fluctuations can occur and hence the
parameter $\epsilon$ should vanish as well.

On the other hand, the two systems yield two different expressions for the sum 
of the matrix elements of the residual effective interaction squared, namely
\label{eq:belu}
\begin{eqnarray}
  \label{eq:b2l}
  \beta^2_l &=& \sqrt{{\cal N}_2/2} \frac{E_l-\bar{\epsilon}_2}{2} 
    \left\{ \left[E_l(1+\sqrt{{\cal N}_2/2})
    -(\epsilon+\sqrt{{\cal N}_2/2}\bar{\epsilon}_2)-E_{\rm HF}\right]
    \right. \\ 
  && \left.
    + \sqrt{ \left[E_l(1+\sqrt{{\cal N}_2/2})
    -(\epsilon+\sqrt{{\cal N}_2/2}\bar{\epsilon}_2)
    - E_{\rm HF}\right]^2 + 4\epsilon(E_l-E_{\rm HF})} \right\} \nonumber
\end{eqnarray}
and
\begin{eqnarray}
  \label{eq:b2u}
  \beta^2_u &=& \sqrt{{\cal N}_2/2} \frac{E_u-\bar{\epsilon}_2}{2} 
    \left\{ \left[E_u(-1+\sqrt{{\cal N}_2/2})
    -(-\epsilon+\sqrt{{\cal N}_2/2}\bar{\epsilon}_2)+E_{\rm HF} 
    \right] \right. \\
  && \left.
    +\sqrt{\left[E_u(-1+\sqrt{{\cal N}_2/2})
    -(-\epsilon+\sqrt{{\cal N}_2/2}\bar{\epsilon}_2)
    + E_{\rm HF}\right]^2 + 4\epsilon(E_u-E_{\rm HF})} \right\} , \nonumber
\end{eqnarray}
both of which vanish in the limit $\epsilon\to0$, in accord with the previous
discussion.

Formula (\ref{eq:b2l}), solution of the first system of equations (the one 
with the ``+'' sign on the right hand side of (\ref{eq:fluctone})), yields the 
value of the {\em energy dependent} quantity (\ref{eq:beta2}) (here denoted by 
$\beta^2_l$) on the lower border of the energy band expressing the fluctuations
of the ground state energy $E$ (and thus encompassing $\bar{E}_0$).
Formula (\ref{eq:b2u}), solution of the second system of equations (the one
with the minus sign  on the right hand side of (\ref{eq:fluctone})), provides
instead (\ref{eq:beta2}) (here denoted by $\beta^2_u$) on the upper border of
the band (remember that $E-\bar{\epsilon}_2<0$).

Of course, of the energy per particle $E$ we only know the experimental value,
not the values on the borders of the band: As a consequence, we can only 
surmise the width $W$ of the latter, thus
providing two different inputs for the energy $E$ appearing on the right hand
side of (\ref{eq:belu}), namely $E_u=E+W/2$ and $E_l=E-W/2$.
However, we can explore whether, for a given $W$, a
value for the parameter $\epsilon$ can be found (not too large, not too small)
such to have the two mean field energies per particle to coincide.
If this search succeeds, then an orientation on $W$ (or, equivalently,
on the size of the fluctuations of the ground state energy) can be gained.

Note that the framework above outlined holds because the same quantity
$\beta^2$ appears in both (\ref{eq:E0barone}) and (\ref{eq:fluctone}). This
occurrence stems from an approximation whose validity is discussed in
Ref.~\cite{DeP99}. 

We now extend the formalism by letting the projector $P$ to 
encompass, beyond the ground state, the 2p-2h excitations of the HF variational
scheme as well. Thus, instead of Eq.~(\ref{eq:Pone}), we write
\begin{equation}
  \label{eq:P2}
  P = |\chi_{\rm HF}\rangle\langle\chi_{\rm HF}| + 
  \sum_\beta|\chi_{\rm HF}^{2\beta}\rangle\langle\chi_{\rm HF}^{2\beta}|,
\end{equation}
where the sum is meant to be extended to the whole set of 2p-2h HF excitations
$|\chi_{\rm HF}^{2\beta}\rangle$.

With the choice (\ref{eq:P2}) the mean field Hamiltonian (\ref{eq:Hbarav}) is
then defined and one can compute the mean field ground state energy per
particle, $\bar{E}_0 = \langle\phi_0|\bar{\cal H}|\phi_0\rangle$, using for the
ket $|\phi_0\rangle$ the expression
\begin{equation}
  \label{eq:phi02}
  |\phi_0\rangle = s_0|\chi_{\rm HF}\rangle + 
    \sum_{\gamma}s_2^\gamma|\chi_{\rm HF}^{2\gamma}\rangle,
\end{equation}
and accounting for the influence of the $Q$-space on the ground state mean
energy per particle in first order of the complexity expansion, i.~e. by
setting 
\begin{equation}
  \label{eq:Q4}
  Q = \sum_{\gamma} 
    |\chi_{\rm HF}^{4\gamma}\rangle\langle\chi_{\rm HF}^{4\gamma}|,
\end{equation}
the sum running over the whole set of the HF 4p-4h excitations.
In (\ref{eq:phi02}) $s_0$ and $s_2^\gamma$ are complex coefficients, fixed,
in principle, by Eq.~(\ref{eq:Hbareq}) and satisfying the normalization 
condition
\begin{equation}
  |s_0|^2+\sum_{\gamma}|s_2^\gamma|^2 = 1.
\end{equation}
After straightforward, but lengthy, algebra using (\ref{eq:phi02}) and
(\ref{eq:Q4}) one arrives at the following new mean field equation
\begin{eqnarray}
  \label{eq:E0bartwoA}
  \bar{E}_0 &=& |s_0|^2 E_{\rm HF} + 
    \sum_{\gamma}|s_2^\gamma|^2
    \langle\chi_{\rm HF}^{2\gamma}|H|\chi_{\rm HF}^{2\gamma}\rangle +
    2s_0^*\sum_{\gamma}s_2^\gamma
    \langle\chi_{\rm HF}|{\cal V}|\chi_{\rm HF}^{2\gamma}\rangle +
    \nonumber \\
  && +\frac{1}{\bar{E}_0-\epsilon-E}\sum_{\beta\gamma}|s_2^\gamma|^2
    |\langle\chi_{\rm HF}^{4\beta}|V|\chi_{\rm HF}^{2\gamma}\rangle|^2,
\end{eqnarray}
${\cal V}$ being the bare NN potential.

Notably, Eq.~(\ref{eq:E0bartwoA}) turns out to formally coincide with
(\ref{eq:E0barone}). Indeed, the first three terms on the right hand side of
Eq.~(\ref{eq:E0bartwoA}) just yield the mean value of the original bare
Hamiltonian $H$ in the state (\ref{eq:phi02}).
In the thermodynamic limit of nuclear matter out of these pieces 
only the HF energy survives. 
Indeed, the correction to the HF energy per particle due to a finite number of
particle-hole excitations vanishes in the thermodynamic limit. In other 
words, the 2p-2h admixture into (\ref{eq:phi02}) {\em does not change} the 
expectation value of the Hamiltonian.

Hence, defining 
\begin{equation}
  \label{eq:zeta2}
  \zeta^2 = \sum_{\beta\gamma} |s_2^\gamma|^2
    |\langle\chi_{\rm HF}^{4\beta}|V|\chi_{\rm HF}^{2\gamma}\rangle|^2,
\end{equation}
(\ref{eq:E0bartwoA}) can be recast into the form
\begin{equation}
  \label{eq:E0bartwo}
  \bar{E}_0 = E_{\rm HF} + \frac{\zeta^2}{\bar{E}_0-\epsilon-E},
\end{equation}
whose similarity with Eq.~(\ref{eq:E0barone}) is transparent.

The same will take place for {\em any} admixture of $N$p-$N$h HF excited states
in $|\chi_{\rm HF}\rangle$: Hence, in our framework different choices of the
projection operator $P$ lead to the same structure for the mean field equation
for nuclear matter. This invariance does not hold in finite nuclei.

The only, of course important, difference between (\ref{eq:beta2}) and 
(\ref{eq:zeta2}) relates to the residual interaction $V$, which in 
(\ref{eq:zeta2}) induces transitions from 2p-2h to 4p-4h states, rather than
from the Fermi sphere to the 2p-2h states.

Concerning the statistical fluctuation equation one can again use
(\ref{eq:E-E0}), 
with the state $|\phi_0\rangle$ given now by Eq.~(\ref{eq:phi02}).
Then invoking {\em the randomness of the phases of the $Q$-space wave functions
(RPA)} and proceeding exactly as done in Refs.~\cite{Car98,DeP99}, one
deduces the new fluctuation equation 
\begin{equation}
  \label{eq:flucttwo}
  E-\bar{E}_0 = \pm\frac{1}{E-\bar{\epsilon}_4}\sqrt{\frac{2}{{\cal N}_4}}
  \zeta^2, 
\end{equation}
where $\bar{\epsilon}_4$ denotes the average energy per particle of the 4p-4h 
HF states.
In (\ref{eq:flucttwo}), ${\cal N}_4$ represents the number of 4p-4h excitations
contributing to the sum over the index $\beta$ in (\ref{eq:zeta2}).

Hence, the ``formal invariance'', with
respect to the choice for the projector $P$, holds for both the equations at
the core of our statistical approach to nuclear matter in first order of the
complexity expansion.

Indeed, the inclusion of $N$p-$N$h states (with $N>2$) into the $P$-space would
merely imply the replacement, in (\ref{eq:flucttwo}), of ${\cal N}_4$ with 
${\cal N}_{N+2}$ and, at the same time, to have $\zeta^2$ defined in terms of
the matrix elements of $V$ between $N$p-$N$h and $(N+2)$p-$(N+2)$h states.
In addition, one should of course insert in the energy denominator the average
energy per particle of the $(N+2)$p-$(N+2)$h HF states.

Therefore, the extension of the $P$-space rapidly leads to the vanishing of the
fluctuations, owing to the very fast increase of the number ${\cal N}_N$.

However, as already mentioned, Eq.~(\ref{eq:E-E0}) {\em is 
not} in general a good starting point to derive the fluctuation equation.
Indeed, it selects out only {\em one term} in the spectral decomposition of the
operator $1/(E-\bar{\cal H})$, which is appropriate for a one-dimensional
$P$-space only. Hence, in place of (\ref{eq:spect}), we rather write
\begin{equation}
  \label{eq:spect2}
  |P\psi\rangle = 
    |\phi_0\rangle\frac{\langle\phi_0|V_{PQ}|Q\psi\rangle}{E-\bar{E}_0} +
    |\phi_2\rangle\frac{\langle\phi_2|V_{PQ}|Q\psi\rangle}{E-\bar{E}_2} +
    \left(\frac{1}{E-\bar{\cal H}}\right)^{\prime\prime} V_{PQ}|Q\psi\rangle,
\end{equation}
with an obvious meaning of the double primed operator in the last term on the
right hand side.
Clearly, Eq.~(\ref{eq:spect2}) does not follows directly from (\ref{eq:P2}),
but it assumes that in the spectral decomposition of the
$(E-\bar{\cal H})^{-1}$ operator only one prominent (collective) state
$|\phi_2\rangle$ enters beyond the ground state $|\phi_0\rangle$.

Sticking to this model, instead of Eq.~(\ref{eq:Qpsipsi}), we likewise write
\begin{equation}
  \label{eq:Qpsipsi2}
  |Q\psi\rangle = \frac{1}{E-h^{(2)}_{QQ}}V_{QP}\left\{
    |\phi_0\rangle\langle\phi_0|P\psi\rangle +
    |\phi_2\rangle\langle\phi_2|P\psi\rangle\right\},
\end{equation}
being
\begin{equation}
  \label{eq:hQQ2}
  h^{(2)}_{QQ} = H_{QQ} + 
    V_{QP}\left(\frac{1}{E-\bar{\cal H}}\right)^{\prime\prime}V_{PQ}.
\end{equation}
Then, by left multiplying (\ref{eq:spect2}) (with $E=E_0$) by $\langle\phi_0|$ 
and using (\ref{eq:Qpsipsi2}), we obtain
\begin{equation}
  \label{eq:E-E02}
  E_0-\bar{E}_0 = 
    \langle\phi_0|V_{PQ}\frac{1}{E_0-h^{(2)}_{QQ}}V_{QP}|\phi_0\rangle +
    \langle\phi_0|V_{PQ}\frac{1}{E_0-h^{(2)}_{QQ}}V_{QP}|\phi_2\rangle
    \frac{\langle\phi_2|P\psi\rangle}{\langle\phi_0|P\psi\rangle},
\end{equation}
which generalizes Eq.~(\ref{eq:E-E0}).

In a perfectly analogous fashion, by left multiplying (\ref{eq:spect2}) (with 
$E=E_2$) by $\langle\phi_2|$ and using (\ref{eq:Qpsipsi2}), we obtain
\begin{equation}
  \label{eq:E-E2}
  E_2-\bar{E}_2 = 
    \langle\phi_2|V_{PQ}\frac{1}{E_2-h^{(2)}_{QQ}}V_{QP}|\phi_2\rangle +
    \langle\phi_2|V_{PQ}\frac{1}{E_2-h^{(2)}_{QQ}}V_{QP}|\phi_0\rangle
    \frac{\langle\phi_0|P\psi\rangle}{\langle\phi_2|P\psi\rangle}.
\end{equation}
In the above, $E_0$ and $E_2$ stand for the first two exact eigenvalues
of the Schroedinger equation; $\bar{E}_0$ and $\bar{E}_2$ for the corresponding
quantities associated with Eq.~(\ref{eq:Hbareq}). 
Eq.~(\ref{eq:E-E2}) shows that, in the present framework, all the energies of
the $P$-space fluctuate. 

Now, the energy averaging of (\ref{eq:E-E02}) and (\ref{eq:E-E2}) vanishes by
definition, but the energy averaging of their square, which yields the
``error'', does not. Hence, proceeding along the lines of
Refs.~\cite{Car98,DeP99}, we subtract on the right hand side of both equations
their average values, square the expressions thus obtained and make use of RPA,
keeping of our expansion in the complexity of the $Q$-space states the first
term only.
Next, we exploit the structure of $|\phi_2\rangle$, which, like
$|\phi_0\rangle$, must be normalized, orthogonal to $|\phi_0\rangle$ and of the
form
\begin{equation}
  \label{eq:phi2HF}
  |\phi_2\rangle = \sum_{\beta}c_2^\beta|\chi_{\rm HF}^{2\beta}\rangle +
    c_0|\chi_{\rm HF}\rangle .
\end{equation}
Finally, we arrive at the equation
\begin{equation}
  \label{eq:flucttwo0}
  E_0-\bar{E}_0 = \pm\frac{1}{E_0-\bar{\epsilon}_4}\sqrt{\frac{2}{{\cal N}_4}}
  (\zeta^2+r\xi^2), 
\end{equation}
where, in addition to (\ref{eq:zeta2}), the further definition 
\begin{equation}
  \label{eq:xi2}
  \xi^2 = \sum_{\beta\gamma} {s_2^\gamma}^* c_2^\gamma
    |\langle\chi_{\rm HF}^{4\beta}|V|\chi_{\rm HF}^{2\gamma}\rangle|^2
\end{equation}
has been introduced; moreover, we have set 
\begin{equation}
  \label{eq:r}
  r \equiv \frac{\langle\phi_2|P\psi\rangle}{\langle\phi_0|P\psi\rangle}.
\end{equation}
Likewise, for the energy of the 2p-2h state of the $P$-space one obtains the
fluctuation equation
\begin{equation}
  \label{eq:flucttwo2}
  E_2-\bar{E}_2 = \pm\frac{1}{E_2-\bar{\epsilon}_4}\sqrt{\frac{2}{{\cal N}_4}}
  (\eta^2+\frac{\xi^2}{r}), 
\end{equation}
where, naturally,
\begin{equation}
  \label{eq:eta2}
  \eta^2 = \sum_{\beta\gamma} |c_2^\gamma|^2
    |\langle\chi_{\rm HF}^{4\beta}|V|\chi_{\rm HF}^{2\gamma}\rangle|^2.
\end{equation}
We thus see that the statistical fluctuation equations (\ref{eq:flucttwo0}) and
(\ref{eq:flucttwo2}) are actually coupled through the term (\ref{eq:xi2}).

Concerning the mean field equations, clearly with the projector (\ref{eq:P2})
an equation should exist also for the energy of the 2p-2h state. It can be
derived by computing $\bar{E}_2=\langle\phi_2|\bar{\cal H}|\phi_2\rangle$ and,
notably, it turns out to read 
\begin{equation}
  \label{eq:E2bar}
  \bar{E}_2 = E^{(2)}_{\rm HF} + \frac{\eta^2}{\bar{E}_2-\epsilon-E_2},
\end{equation}
$E^{(2)}_{\rm HF}$ representing the HF energy per particle of the system
in the 2p-2h excited state.
Since we split $E^{(2)}_{\rm HF}$ 
into a part associated with the HF ground state and a part
associated with the 2p-2h {\em excitation energies}, both per particle, and
since the latter vanishes in the thermodynamic limit, --- as previously noted
in commenting Eq.~(\ref{eq:E0bartwoA}), then
$E^{(2)}_{\rm HF}=E_{\rm HF}$ , a relation we expect to be approximately
fulfilled also in a heavy nucleus.

We conclude from the above analysis that our approach leads to a set of mean 
field equations, one for each of the states lying in the $P$-space: These 
equations, unlike the fluctuation ones, are not coupled.

\subsection{ Normalization and fluctuation of the $P$-space ground state wave 
function 
}
\label{subsec:S}

In the present framework the ground state spectroscopic factor $S$ is the
square root of the norm of $|P\psi\rangle$, the system's ground state wave
function projection in $P$-space. To find it we exploit the completeness of the
normalized eigenstates of $\bar{\cal H}$. Hence we write
\begin{equation}
  \label{eq:Sdef}
  S^2\equiv\langle P\psi|P\psi\rangle
    =\sum_{n=0}^M\langle P\psi|\phi_n\rangle\langle\phi_n|P\psi\rangle
    =1-\langle Q\psi|Q\psi\rangle.
\end{equation}
Now, confining ourselves to set $M=1$, then the Eq.~(\ref{eq:Qpsipsi}) for
$|Q\psi\rangle$ is warranted and we rewrite (\ref{eq:Sdef}) as follows
\begin{equation}
  \label{eq:Sint1}
  S^2 = 1-\langle\phi_0|V_{PQ}\frac{1}{(E_0-h_{QQ})^2}V_{QP}|\phi_0\rangle
    |\langle\phi_0|P\psi\rangle|^2.
\end{equation}
Moreover, when $M=1$ then 
\begin{equation}
  S^2 = |\langle\phi_0|P\psi\rangle|^2.
\end{equation}
Hence, by exploiting (\ref{eq:E-E0}), Eq.~(\ref{eq:Sint1}) can be recast into
the form\cite{Car98,DeP99}
\begin{equation}
  \label{eq:Sint2}
  S^2 = 1+S^2\left[\frac{d}{dE_0}(E_0-\bar{E}_0)
    +\frac{E_0-\bar{E}_0}{\bar{E}_0-\epsilon-E_0}\right].
\end{equation}
Finally, employing (\ref{eq:E0barsol}) the expression 
\begin{equation}
  \label{eq:Sfin}
  S^2 = \left[\frac{3}{2}+\frac{1}{2}
    \frac{E_{\rm HF}-E_0-\epsilon-2d\beta^2/dE_0}
    {\sqrt{(E_{\rm HF}-E_0-\epsilon)^2+4\beta^2}}
    +\frac{\epsilon}{\bar{E}_0-\epsilon-E_0}\right]^{-1}
\end{equation}
follows\cite{Car98,DeP99}, where the energy derivative of the sum of the square
moduls of the vacuum--2p-2h matrix elements of the effective interaction
appears (its explicit expression is given in Ref.~\cite{DeP99}). Note that
(\ref{eq:Sfin}) goes to one as $\beta^2\to0$, as it should.

If, however, the expression for the $Q$-space wave function appropriate for a
two-dimensional $P$-space, namely (\ref{eq:Qpsipsi2}), is used, then (see
Ref.~\cite{DeP02} for details) one ends up with the expression
\begin{equation}
  \label{eq:Sint22}
  S^2 = \frac{1-\left[
    \frac{\displaystyle d(\bar{E}_2-\bar{E}_0)}{\displaystyle dE_0}+
    \frac{\displaystyle\bar{E}_2-\bar{E}_0}
    {\displaystyle\bar{E}_0-\epsilon-E_0}\right]
    |\langle\phi_2|P\psi\rangle|^2}
    {1+\frac{\displaystyle\epsilon}{\displaystyle\bar{E}_0-\epsilon-E_0}
    +\frac{\displaystyle d\bar{E}_0}{\displaystyle dE_0}},
\end{equation}
which reduces to (\ref{eq:Sint2}), as it should, if
$\langle\phi_2|P\psi\rangle\to0$, i.~e. for a one-dimensional $P$-space.
In deducing (\ref{eq:Sint22}) the approximation 
\begin{equation}
  \frac{1}{E_0-h_{QQ}^{(2)}} \approx \frac{1}{E_2-h_{QQ}^{(2)}}
\end{equation}
has been made.

Since, from Eq.~(\ref{eq:r}),
\begin{equation}
  \label{eq:phi2psi}
  |\langle\phi_2|P\psi\rangle|^2 = \frac{r^2}{1+r^2}S^2,
\end{equation}
then Eq.~(\ref{eq:Sint22}) can be recast as follows
\begin{equation}
  \label{eq:Sdef2}
  S^2 = \left\{1+\frac{1}{\bar{E}_0-\epsilon-E_0}\left[\epsilon+
    \frac{r^2}{1+r^2}\left(\bar{E}_2-\bar{E}_0\right)\right]+
    \frac{1}{1+r^2}\left(
    \frac{d\bar{E}_0}{dE_0}+r^2\frac{d\bar{E}_2}{dE_0}\right)\right\}^{-1},
\end{equation}
which again reduces to (\ref{eq:Sint2}) as $r\to0$.

We now address the problem of the fluctuation of $|P\psi\rangle$.
For this scope, we focus on the ground state and, by combining 
Eqs.~(\ref{eq:Hbar0eq}), (\ref{eq:Schrava}) and (\ref{eq:Schravb}), we obtain
\begin{equation}
  (E-\bar{\cal H})[|P\psi\rangle-|\big\langle P\psi\big\rangle\rangle]
    +(E-\bar{E}_0)|\big\langle P\psi\big\rangle\rangle =
    V_{PQ}|Q\psi\rangle
\end{equation}
(the angle brackets meaning energy averaging).

Then, if use is made of the expression (\ref{eq:Qpsipsi}) for $|Q\psi\rangle$ 
and of the spectral decomposition of the operator $(E-\bar{\cal H})^{-1}$, one 
gets 
\begin{eqnarray}
  \label{eq:fluct1}
  |P\psi\rangle-|\big\langle P\psi\big\rangle\rangle &=& 
    {\sum}^\prime|\phi_n\rangle\frac{1}{E-\bar{E}_n}\langle\phi_n|V_{PQ}
    \frac{1}{E-h_{QQ}}V_{QP}|\phi_0\rangle\langle\phi_0|P\psi\rangle
    \nonumber \\
  && + \frac{1}{E-\bar{E}_0}|\phi_0\rangle\langle\phi_0|V_{PQ}
    \frac{1}{E-h_{QQ}}V_{QP}|\phi_0\rangle\langle\phi_0|P\psi\rangle
    \nonumber \\
  && - (E-\bar{E}_0)\sum_n\frac{1}{E-\bar{E}_n}|\phi_n\rangle\langle\phi_n|
    \big\langle P\psi\big\rangle\rangle.
\end{eqnarray}
Now, since 
$|\big\langle P\psi\big\rangle\rangle\propto|\phi_0\rangle$, from the above
finally it follows 
\begin{eqnarray}
  \label{eq:fluct2}
  |P\psi\rangle-|\big\langle P\psi\big\rangle\rangle &=& 
    (1-|\phi_0\rangle\langle\phi_0|)^{-1}
    \left(\frac{1}{E-\bar{\cal H}}\right)^\prime V_{PQ}\frac{1}{E-h_{QQ}}
    V_{QP}|\phi_0\rangle\langle\phi_0|P\psi\rangle \nonumber \\
  &=& \langle\phi_0|P\psi\rangle
    \left(\frac{1}{E_0-\bar{\cal H}}\right)^\prime 
    V_{PQ}\frac{1}{E_0-h_{QQ}}V_{QP}
    |\phi_0\rangle,
\end{eqnarray}
which {\em vanishes when $P$ is given by (\ref{eq:Pone})}, since clearly 
$|P\psi\rangle$ does not fluctuate in a one-dimensional $P$-space.

If, on the other hand, $P$ is given by Eq.~(\ref{eq:P2}), then
the above can be computed in first order of the complexity expansion, using
(\ref{eq:phi02}) for $|\phi_0\rangle$, (\ref{eq:phi2HF})
for $|\phi_2\rangle$ and (\ref{eq:Q4}) for $Q$. One ends up with the expression
(we set $E=E_0$ to conform to previous notations) 
\begin{equation}
  [|P\psi\rangle-|\big\langle P\psi\big\rangle\rangle]_1 = 
    \langle\phi_0|P\psi\rangle\frac{|\phi_2\rangle}{E_0-\bar{E}_2}
    \sum_{\beta\beta'\gamma}
    \langle\chi_{\rm HF}^{2\beta}|V|\chi_{\rm HF}^{4\gamma}\rangle
    \frac{c^*_{2\beta}s^{\phantom{*}}_{2\beta'}}
    {E_0-\epsilon_{\rm HF}^{4\gamma}}
    \langle\chi_{\rm HF}^{4\gamma}|V|\chi_{\rm HF}^{2\beta'}\rangle,
\end{equation}
which can be further simplified invoking the randomness of the phases of the
wave functions in the $Q$-space and again introducing the average energy per
particle $\bar{\epsilon}_4$ for the 4p-4h HF excited states. 
Then, with the help of Eq.~(\ref{eq:xi2}), the formula
\begin{equation}
  \label{eq:psiflucttwo}
  [|P\psi\rangle-|\big\langle P\psi\big\rangle\rangle]_1 =
    \frac{\langle\phi_0|P\psi\rangle}
    {(E_0-\bar{E}_2)(E_0-\bar{\epsilon}_4)}|\phi_2\rangle\xi^2
\end{equation}
is derived. It gives the fluctuations of the wave function in first order of 
the complexity expansion.

Notice that in (\ref{eq:psiflucttwo}) the scalar product $\langle\phi_0|P\psi\rangle$, unlike in 
Refs.~\cite{Car98,DeP99}, does not coincide with the spectroscopic factor
$S$, as defined in (\ref{eq:Sdef}), {since Eq.~(\ref{eq:psiflucttwo}) refers to
a $P$-space with dimensions larger than one. Rather, it measures the amount of
the true ground state wave function of the system embodied in the mean field 
state $|\phi_0\rangle$.

The relevance of Eq.~(\ref{eq:psiflucttwo}) lies in the possibility it offers
to assess the ``error'' affecting the ground state expectation value of
operators associated to physical observables like, e.~g., the magnetic moments
of nuclei.

\section{A glance at QCD}
\label{sec:QCD}

Lately the existence of randomness has been unraveled also at the scale of
quarks, namely at the level of the physics ruled by QCD. Here the RMT has
achieved outstanding successes in connection with the low-lying eigenvalues of
the Dirac operator for massless fermions:
\begin{equation}
\label{eq:Dslash}
  i\rlap/D = i\rlap/\partial + g \sum_a\frac{\lambda^a}{2} \rlap/A^a ,
\end{equation}
where $g$ is the coupling constant, the matrices $\lambda^a$ the generators of
the gauge group and the $\rlap/A^a$ the gauge fields.
Indeed, the spectrum of (\ref{eq:Dslash}),
\begin{equation}
\label{eq:iDslashphin}
  i\rlap/D \phi_n = \lambda_n \phi_n ,
\end{equation}
has been computed for a SU(2) gauge theory (and for others as well) and from
the $\lambda_n$'s the nearest-neighbor spacing distribution $P(s)$ has been
deduced. In turn, this has been compared to the prediction of the chRMT (chiral
random matrix theory): An impressive agreement has been found,\cite{Hal97}
pointing to the universality of the fluctuations of the eigenvalues of the
Dirac operator.

The focus on the $\lambda_n$'s also stems from the anticommutator
\begin{equation}
  \{i\rlap/D,\gamma_5\} = 0,
\end{equation}
whose vanishing in the chiral limit implies for the eigenvalues of
(\ref{eq:iDslashphin}) the occurrence in symmetrical pairs ($\pm\lambda_n$)
around $\lambda=0$. As a consequence the $\lambda_n$'s tend to accumulate
around the origin and their average density 
\begin{equation}
  \rho(\lambda) = \langle\sum_n\delta(\lambda-\lambda_n)\rangle,
\end{equation}
the average being taken on all the configurations of the gauge field, on the
one hand displays the level repulsion already encountered at the level of the
physics of the nucleus, --- thus signalling universality, --- and on the other
plays a central role in the physics of the quark-gluon plasma.

In fact, the density of the smallest eigenvalues of the Dirac operator is
directly related to the chiral condensate according to
Banks--Casher formula\cite{Ban80}:
\begin{equation}
  \langle\bar{q} q\rangle = -\pi\rho(0)/V,
\end{equation}
$V$ being the space-time volume.
It is fascinating that such a fundamental aspect of nature as the breaking of
the chiral symmetry appears to be ruled by chaos.

In conclusion, we feel that RMT, in both its normal and chiral versions, is
invaluable in disentangling the stochastic content of nuclear physics and QCD.
In the former case, in the Feshbach's approach, stochasticity should be
averaged out, thus leading to an effective Hamiltonian that has been proved
most successful in interpreting the data. We argue that the same path might be
worth following also in QCD.

\section*{Acknowledgments}

We would like to express our gratitude to H. Weidenmueller for many enlighting
discussions and to W. M. Alberico for the kind invitation to this workshop.

\end{document}